\documentclass{appolb}
\usepackage{epsfig}

\begin{document}
\newcommand {\snn} {$\sqrt{s_{_{NN}}}=200$~GeV}
\newcommand {\pt} {p_T}
\newcommand {\ptt} {p_T^{\rm trig}}
\newcommand {\pta} {p_T^{\rm assoc}}
\newcommand {\ptref} {p_T^{\rm ref}}
\newcommand {\deta} {\Delta\eta}
\newcommand {\dphi} {\Delta\phi}
\newcommand {\vv}[2] {$v_{#1}\{#2\}$}
\newcommand {\VV}[1] {V_{#1\Delta}}

\pagestyle{plain}
\newcount\eLiNe\eLiNe=\inputlineno\advance\eLiNe by -1
\title{High-$\pt$ triggered dihadron correlations with $v_n$ background subtraction by STAR
}
\author{Fuqiang Wang (for the STAR Collaboration)
\address{Department of Physics, Purdue University, West Lafayette, Indiana 47907, USA}
}
\maketitle

\begin{abstract}
STAR measurements of dihadron correlations with high-$\pt$ trigger partcles in Au+Au collisions at \snn\ are presented with subtraction of $v_2$, $v_3$, and $v_4$ backgrounds. The $v_n$ azimuthal anisotropies were measured by the two-particle cumulant method with $\eta$ gap to reduce nonflow. The dihadron correlations relative to the $v_2$ as well as the $v_3$ harmonic planes are also presented. Implications of the results are discussed. 
\end{abstract}

\section{Introduction}
%
%
Relativistic heavy ion collisions create a hot and dense medium exhibiting hydrodynamic properties. The particle azimuthal distributions are anisotropic resulting from hydrodynamic conversion of the initial spatial anisotropy. Hydrodynamics describe well the expansion and anisotropies of particles at transverse momentum $\pt<2$~GeV/$c$.
Hard-scattering probes, on the other hand, interact with the medium and lose energy, resulting in depletion of high-$\pt$ hadrons in the final state. This jet-quenching phenomenon provides an exclusive channel to study QCD interactions and deduce the properties of the QCD matter at high energy densities. 

%
Dihadron correlations with high-$\pt$ trigger particles have proven valuable in jet-quenching studies. One of the difficulties in dihadron correlation measurements is the subtraction of combinatoric background. The background is not uniform but modulated by flow harmonics (correlation to the event plane). 
%
%
Triggered correlations have been measured with only $v_2$ and $v_4$ subtraction, assuming vanishing triangular flow ($v_3$). Novel structures were observed constituting a near-side long-range speudo-rapidity ($\deta$)  correlation (`ridge')~\cite{PRL95,Putschke} and a away-side double-peak azimuthal ($\dphi$) correlation (`Mach-cone')~\cite{PRL95,Lacey,PRL102,PRC10}. 


Recent model studies suggest, however, that triangular and other odd harmonics do not vanish because of initial geometry fluctuations~\cite{Alver}. A non-zero $v_3\{2\}$ from two-particle cumulant method has been measured~\cite{ALICE,Sorensen}; 
$v_n\{2\}=\sqrt{\VV{n}}$, $\VV{n}\equiv\langle\cos(n\dphi)\rangle$ where $\dphi$ is the two-particle opening azimuthal angle. While a non-zero $v_3\{2\}$ itself is not a proof of finite hydrodynamic triangular flow because nonflow also contributes, the centrality and $\pt$ dependences of the measured $v_3\{2\}$ do suggest that part of the $v_3\{2\}$ is of hydrodynamic origin, and this part should be subtracted from dihadron correlations. Two questions arise: (1) How much does nonflow (correlation unrelated to the reaction plane) contribute to the measured $v_n\{2\}$? (2) What are the effects of jet-medium interactions?


This talk does not provide a complete answer to these questions, but makes progress towards that end. We report STAR measurements of $v_2$ and $v_3$ in Au+Au collisions at \snn. We subtract  $v_n$ backgrounds and present the obtained dihadron correlations integrated over as well as relative to the $v_2$ and $v_3$ harmonic planes. We discuss possible implications of our results. 

\section{Dihadron correlation with $v_n$ subtraction}

\begin{figure}[hbt]
\vspace*{-0.2in}
\begin{center}
\includegraphics[width=\textwidth]{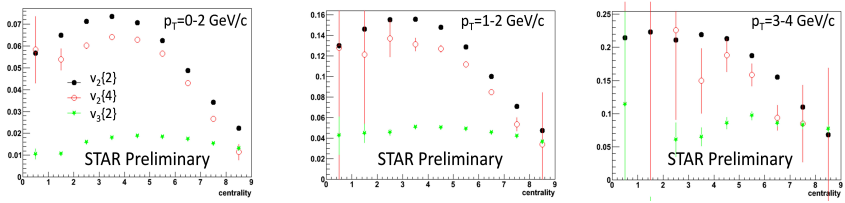}
\end{center}
\vspace*{-2.7in}
\caption{(Color online) Two- and four-particle anisotropy measurements \vv{2}{2}, \vv{2}{4}, \vv{3}{2} as a function of centrality in Au+Au collisions at \snn\  by STAR. The abscissa bins 0-6 stand for 10\%-size centralities from 80\% to 10\%, and 7-8 for 5-10\% and top 5\% centralities. Three $\pt$ ranges are shown, 0-2 GeV/$c$ (left), 1-2 GeV/$c$ (middle), and 3-4 GeV/$c$ (right).}
\label{fig1}
\end{figure}

Figure~\ref{fig1} shows the \vv{2}{2}\ and \vv{3}{2}\ measurements by the two-particle cumulant method~\cite{Sorensen}. 
The four-particle cumulant \vv{2}{4}\ is also shown for comparison. 
The \vv{3}{2}\ has a weak centrality dependence, consistent with its fluctuation origin. At large $\pt$ in central collisions, $v_3$ is comparable to $v_2$.
An $\eta$-gap, $|\deta|>1$, was applied in the analysis to reduce the nonflow contributions from small-angle correlations, such as jet-like correlations, resonance decays, etc. Nonflow correlations beyond $|\deta|>1$ still remain. One likely contribution comes from away-side jet-like correlations because the away-side jet partner is uncorrelated to the near-side jet in $\eta$. 


Two-particle cumulant \vv{n}{2}\ measures the net effect of flow, flow fluctuations, and nonflow. Except the nonflow contamination, \vv{n}{2}\ is the most truthful background to dihadron correlation. However, if \vv{n}{2}\ is measured from the same pairs used in dihadron correlation analysis, then \vv{n}{2}\ will precisely describe the correlation function and the \vv{n}{2}-subtracted signal will by definition be zero. It is thus important to measure \vv{n}{2}\ using pairs far removed in phase-space from those used in dihadron correlation analysis. 

Dihadron correlations at low-to-intermediate $\pt$ with $\deta$ gap in Pb+Pb collisions at LHC were shown to be completely described by the sum of a few Fourier harmonics~\cite{ALICE}. 
The particle anisotropies at a given $\pt$ were measured by the Fourier coefficients of their correlations to reference particles from $0.2<\ptref<5$~GeV/$c$ with $|\deta|>1$ in ALICE~\cite{ALICE}: $v_n(\pt)=\VV{n}(\pt,\ptref)/\sqrt{\VV{n}(\ptref,\ptref)}$. The dihadron correlation result of Ref.~\cite{ALICE}, with trigger $\ptt$=2-3~GeV/$c$ and associated $\pta$=1-2~GeV/$c$, thus indicates $\VV{n}(\ptt,\ptref)\VV{n}(\pta,\ptref)\approx\VV{n}(\ptt,\pta)\VV{n}(\ptref,\ptref)$, which is not surprising given the relatively small nonflow contributions (as gauged by the \vv{2}{2} and \vv{2}{4} measurements). Quantitatively how much nonflow contributes to the measured $\VV{n}$ will require further studies. 

\begin{figure}[hbt]
\vspace*{-0.2in}
\begin{center}
\includegraphics[width=\textwidth]{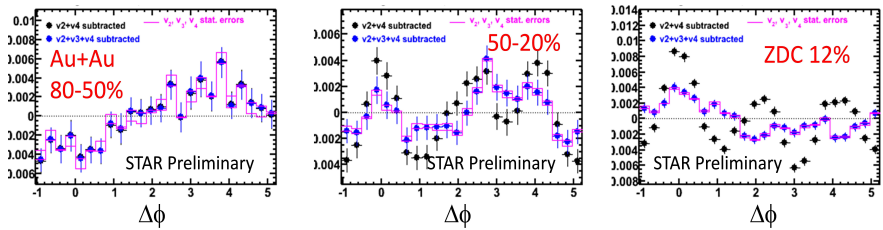}
\end{center}
\vspace*{-2.6in}
\caption{(Color online) $v_n$-subtracted dihadron correlations at $|\deta|>1$ in Au+Au collisions at \snn\ by STAR. Three centrality ranges are shown: 50-80\% (left), 20-50\% (middle), and ZDC top 12\% (right). The trigger and associated $\pt$ ranges are 3-6 GeV/$c$ and 2-3 GeV/$c$, respectively. The subtracted \vv{n}{2}\ were measured by the two-particle cumulant method with $|\deta|>1$ and reference particle $\ptref<2$~GeV/$c$. The background is normalized such that the average signal is zero.}
\label{fig2}
\end{figure}

Figure~\ref{fig2} shows the dihadron correlation signals in Au+Au collisions at \snn\ by STAR after $v_n$ subtraction. The trigger and associated $\pt$ ranges are 3-6~GeV/$c$ and 2-3~GeV/$c$, respectively. The \vv{n}{2}\ were measured with reference particles from $\ptref<2$~GeV/$c$, which is somewhat removed from the trigger and associated $\pt$ regions. The blue data points show the results with $v_2$, $v_3$, and $v_4$ subtraction, while the black ones show those with only $v_2$ and $v_4$ subtraction for comparison. In peripheral collisions, no visible signal remains besides a negative dipole. This may suggest that the subtracted $v_n$ contain nonflow contributions comparable to the jet-like correlation between the trigger-associated pairs. In medium-central collisions, there seems finite correlation amplitude on the near side which may indicate a ridge contribution. In other words, the previously observed ridge may not be completely explained by $v_3$. On the away side, a broad correlation signal is still observed, however, the double-hump structure is weaker. In the most central collisions, no away-side correlation is visible, consistent with the ``disapparence'' of away-side high-$\pt$ hadrons. The near-side correlation seems to indicate a finite peak consistent with remaining ridge contributions.

\section{Dihadron correlation relative to the event plane}

\begin{figure}
\vspace*{-0.2in}
\begin{center}
\includegraphics[width=\textwidth]{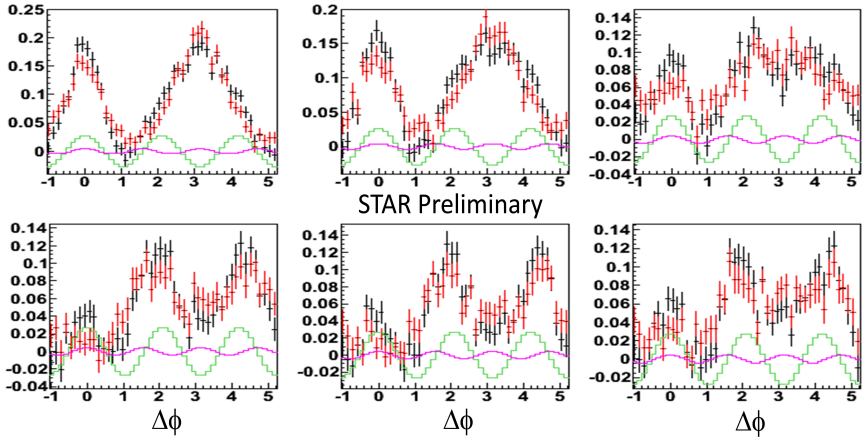}
\end{center}
\vspace*{-1.4in}
\caption{(Color online) $v_n$-subtracted dihadron correlations at $|\deta|>0.7$ in 20-60\% Au+Au collisions at \snn\ by STAR. The correlations are shown, from upper left (in-plane) to lower right (out-of-plane) panel, in $15^\circ$ steps in trigger particle azimuth relative to the event plane, $\phi_s$. The trigger and associated $\pt$ ranges are 3-4 GeV/$c$ and 1-2 GeV/$c$, respectively. The subtracted $v_n$ were measured by the two-particle cumulant method with $|\deta|>0.7$ and reference particle $\ptref<2$~GeV/$c$. The subtracted $v_3$ background (uncorrelated to the event plane) and the uncorrelated portion of $v_4$ are shown in the green and pink histograms, respectively. The background is normalized by the ZYAM prescription.}
\label{fig3}
\end{figure}

Dihadron correlations as a function of the trigger particle azimuth relative to the event plane, $\phi_s=\phi_t-\psi_2$, have been analyzed in 20-60\% Au+Au collisions by STAR~\cite{corrEP}. The $v_2$ and $v_4$ backgrounds were subtracted. The $v_2$ was obtained from the average of two- and four-particle cumulant methods~\cite{PRC75}. The $v_4$ was measured with respect to the $v_2$ harmonic event plane~\cite{PRC75}. The effect of $v_3$ background was estimated and found to be insignificant~\cite{corrEP}. In this talk we use for background subtraction the measured \vv{2}{2}\ ($|\deta|>0.7$) and the parameterized $v_4\{\psi_2\}=1.15v_2\{2\}^2$ for $v_4$ correlated to $\psi_2$. The resultant dihadron correlations at $|\deta|>0.7$ are shown in the black histograms in Fig.~\ref{fig3} and are consistent with the results in Ref.~\cite{corrEP}. We further subtract the additional \vv{3}{2}\ and $v_4\{{\rm uncorr.}\}=\sqrt{v_4\{2\}^2-v_4\{\psi_2\}^2}$ backgrounds (both uncorrelated to $\psi_2$):
$2v_3^{\rm trig}\{2\}v_3^{\rm assoc}\{2\}\cos3\dphi+2v_4^{\rm trig}\{{\rm uncorr.}\}v_4^{\rm assoc}\{{\rm uncorr.}\}\cos4\dphi$.
The results are shown in Fig.~\ref{fig3} in the red histograms. 
The results again show that the effect of $v_3$ is insignificant; however, the reduction in the near-side correlation magnitude is noticeable.

\begin{figure}
\vspace*{-0.2in}
\begin{center}
\includegraphics[width=\textwidth]{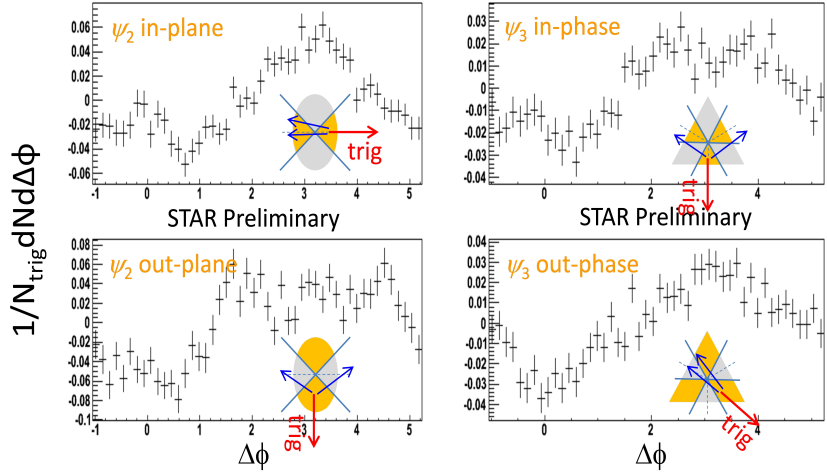}
\end{center}
\vspace*{-1.2in}
\caption{(Color online) $v_n$-subtracted dihadron correlations at $|\deta|>0.7$ in 20-60\% Au+Au collisions at \snn\ by STAR. The correlations are shown for trigger particles in-plane (upper-left) and out-of-plane (lower-left) of the $v_2$ event plane ($\psi_2$), and for trigger particles in-phase (upper-right) and out-of-phase (lower-right) of the $v_3$ harmonic plane ($\psi_3$). The trigger and associated $\pt$ ranges are 3-4 GeV/$c$ and 1-1.5 GeV/$c$, respectively. The subtracted $v_n$ were measured by the two-particle cumulant method with $|\deta|>0.7$ and reference particle $\ptref<2$~GeV/$c$. The background is normalized such that the average signal is zero.}
\label{fig4}
\end{figure}

The event-plane dependent dihadron correlation analysis can be carried out with respect to the $v_3$ harmonic plane $\psi_3$. In reconstructing $\psi_3$ (and $\psi_2$), particles within $|\deta|<0.5$ of the trigger particle are excluded. Analogous to correlations relative to $\psi_2$, the $v_3$ background depends on the trigger particle azimuth $\phi_{s,3}=\phi_t-\psi_3$. The $v_2$ and $v_4$ backgrounds are independent of $\phi_{s,3}$. Figure~\ref{fig4} shows in the right panels the dihadron correlation for trigger particles in-phase of $\psi_3$ ($|\phi_{s,3}|<\pi/6$) and out-of-phase ($|\phi_{s,3}-\pi/3|<\pi/6$). For comparison those for trigger particles in-plane of $\psi_2$ ($|\phi_s|<\pi/4$) and out-of-plane ($|\phi_s-\pi/2|<\pi/4$) are shown in the left panels. The cartoon inserts help visulize the geometry. The correlation results seem to indicate a near-side ridge for trigger particles both in-plane of $\psi_2$ and in-phase of $\psi_3$. For away side, the correlation is broader for triggers out-of-plane than in-plane of $\psi_2$. To the contrary, the correlation is broader for triggers in-phase than out-of-phase of $\psi_3$. These results seem to suggest path-length or geometry effects on the away-side jet-like correlations. However, the exact nature of the effects need further investigations.

\section{Summary}

We have presented dihadron correlations with trigger $\ptt$=3-6~GeV/$c$ and associated $\pta$=2-3~GeV/$c$ in Au+Au collisions at \snn\ by STAR, with $v_2$, $v_3$, and $v_4$ background subtraction. The $v_n$ anisotropy were measured by two-particle cumulant method with reference particles from $0.2<\ptref<2$~GeV/$c$ and with a relatively large $\eta$ gap. A near-side `ridge' peak seems to remain in non-peripheral collisions. The away-side correlation, while strongly suppressed in central collisions, is broad in medium-central collisions. 

We have also presented dihadron correlations relative to the $v_2$ harmonic event plane $\psi_2$ as well as the $v_3$ harmonic plane $\psi_3$, also with $v_2$, $v_3$, and $v_4$ background subtraction, in 20-60\% Au+Au collisions. The effect of $v_3$ is small. A near-side ridge seems to be present in-plane of $\psi_2$ (and in-phase of $\psi_3$), and disappear out-of-plane of $\psi_2$ (and out-of-phase of $\psi_3$). The away-side correlation is single peaked for triggers in-plane of $\psi_2$ and broadens out-of-plane. This trend seems reversed relative to $\psi_3$--The away-side correlation is single-peaked for triggers out-of-phase of $\psi_3$ and broadens in-phase. These results seem to indicate nontrivial path-length or geometry effects of the underlying physics. 

\end{document}